# DeepHider: A Covert NLP Watermarking Framework Based on Multi-task Learning

## Long Dai, Jiarong Mao, Xuefeng Fan and Xiaoyi Zhou*

*School of Cyberspace Security, Hainan University, Haikou 570100, China*

**Abstract**

Natural language processing (NLP) technology has shown great commercial value in applications such as sentiment analysis. But NLP models are vulnerable to the threat of pirated redistribution, damaging the economic interests of model owners. Digital watermarking technology is an effective means to protect the intellectual property rights of NLP model. The existing NLP model protection mainly designs watermarking schemes by improving both security and robustness purposes, however, the security and robustness of these schemes have the following problems, respectively: (1) Watermarks are difficult to defend against fraudulent declaration by adversary and are easily detected and blocked from verification by human or anomaly detector during the verification process. (2) The watermarking model cannot meet multiple robustness requirements at the same time. To solve the above problems, this paper proposes a novel watermarking framework for NLP model based on the over-parameterization of depth model and the multi-task learning theory. Specifically, a covert trigger set is established to realize the perception-free verification of the watermarking model, and a novel auxiliary network is designed to improve the robustness and security of the watermarking model. The proposed framework was evaluated on two benchmark datasets and three mainstream NLP models, and the results show that the framework can successfully validate model ownership with 100% validation accuracy and advanced robustness and security without compromising the host model performance.

## 1. Introduction

In recent years, major Internet manufacturers have launched NLP services and applications with deep neural network as the core. They upload well-trained models to the cloud and provide API for paying customers to query. It is well known that the design of NLP models requires large amounts of annotated data and the expertise of model designers, as well as high performance equipment to support and consume a lot of time for training. As a result, deployed models are vulnerable to theft by profit-driven internal or external adversaries for illegal distribution, sale, or establishment of illegal NLP services for profit, which can seriously threaten the normal business of model owners and their intellectual property rights. Therefore, in order to protect the intellectual property of the NLP model, a method of remotely verifying the ownership of the model is required.

Digital watermarking [1, 2] is a technology of embedding hidden marks in multimedia data by means of signal processing, which is widely used in copyright protection of multimedia content. If multimedia content is misappropriated by illegals, the owner can extract the watermark from the protected multimedia and thus prove its intellectual property rights. Based on such characteristics, in recent years, researchers have extended digital watermarking technology to copyright protection of DNN model [3-12]. DNN model watermarking schemes [3-9] mainly focus on protecting image classification and processing models, and these schemes can be divided into black-box verification [5-9] and white-box verification [3, 4, 9]. Black-box verification is done by adding corresponding perturbations to the images and assigning specific target labels to them, and the verification simply uses the image trigger set to access the interface constructed by the remote model. However, this trigger set is only applicable to copyright protection of image model and cannot be migrated to NLP model. White-box verification, on the other hand, embeds the owner information into the model weights and verifies the ownership by extracting the internal structure and parameters of the suspicious model during verification. Different from black-box verification, most image watermarking schemes based on white-box verification can be extended to NLP model, but white-box verification needs to obtain the structure and parameters of suspicious model in advance, which has strict conditions. Therefore, it is difficult to fully protect the intellectual property of NLP model by using the existing image model watermarking schemes.

At present, the watermarking scheme of NLP model based on black-box verification [10-12] is still in its infancy. By reviewing the literature, trigger sets built by some NLP backdoor attack schemes [13-16] can also be used to protect the intellectual property rights of NLP models. Their mainstream solutions can be divided into two main categories: word-based and sentence-based approaches. The word-based approach selects a specific rare or neutral word to insert into the text as a trigger set, while the sentence-based approach inserts a neutral sentence into the text as a trigger set. However, inserting words or sentences and changing tenses will destroy the coherence of the original text, and these trigger samples are easy to be detected by manual or anomaly detector and prevent verification. The watermarking security of NLP model based on black-box verification largely depends on concealment of the trigger sample. Therefore, it is necessary to design a covert trigger set to avoid human or anomaly detector detection. It should be noted that the trigger set used for NLP model [10-14, 16] does not pay attention to the reliability index of false positive rate. In order to enhance the reliability of watermark verification, the false positive rate of the trigger set on the cleaning model must be considered in the process of designing the trigger set.

In addition to the concealment and false positive rate of trigger set, the robustness of watermarking is also an important index to evaluate the performance of watermarking. Generally, NLP model watermarking based on black-box validation [10-12] takes global fine-tuning or pruning as the criteria for evaluating robustness. In real cases, adversaries may replace the classification layer of the watermark model and fine-tuning the model through a small number of training samples to meet specific classification tasks, while removing the watermark existing in the model. Therefore, the robustness requirement must be taken into account when designing watermarking scheme. In addition, if the adversary attempts to forge a valid watermark to declare the ownership of the stolen model, the ownership verification of the model owner will be ambiguous. This fraudulent declaration is a problem that is not mentioned in the current NLP watermarking scheme based on black-box verification [10-12].

This paper proposes a novel watermarking framework for NLP models, which aims to protect the intellectual property rights of NLP models. The framework first designs a trigger set with no trigger mode to achieve perception-free authentication of remote NLP models and has a low false positive rate for clean models. Second, a new Secondary Authentication Network (SANet) is designed to simultaneously meet multiple robustness requirements and resist fraudulent declaration by adversaries. In addition, SANet's authentication set and the watermark information embedded inside it can clearly associate the owner's identity with the watermark model, reducing the high costs associated with third-party copyright authentication.

In summary, our work has the following three major contributions:

- A complete watermarking framework is provided for NLP model in black-box setting and white-box setting, which solves the security risks of manual and anomaly detector detection.
- A new Secondary Authentication Network is proposed, which can not only meet a variety of robustness requirements, but also effectively resist the adversary's fraudulent declaration.
- A large number of experiments were conducted on two benchmark datasets and three mainstream NLP models. Experiments verify the effectiveness and universality of the proposed framework, as well as its advanced safety and robustness.

The remainder of this paper is as follows. Section 2 briefly summarizes the work related to the framework, Section 3 gives the preliminary, Section 4 details the specific framework of the proposed watermarking scheme, and extensive experiments and analysis are performed in Section 5, and finally the full paper is summarized in Section 6.

## 2. Related Work

This section reviews and summarizes the work related to the framework. To better understand the current state of DNN watermarking schemes, we detail the image model watermarking scheme, the NLP backdoor attack scheme, and the NLP model watermarking scheme mentioned in Section 1.

### 2.1. Image Model Watermarking

As the most basic tasks in computer vision, image classification model and image processing model have the same commercial value as NLP model. In order to protect the intellectual property of image classification models, Uchida et al. [3] made the first attempt to embed watermarks in image classification models by using a parametric regularizer to embed watermarks into the weight parameters of the convolutional layer of the model and successfully verified model ownership by a white-box approach. However, embedding watermark bit into the weights leads to changes in the weight distribution, which can easily be detected and adjusted accordingly by weight variance analysis. In order to reduce the weight changes caused by watermark embedding, Kuribayashi et al. [4] applied watermarking methods based on quantized index modulation (QIM) to the sampled weight values by fine-tuning the fully connected layer weights. However, the verification of these schemes requires obtaining the stolen model weights in order to extract the watermark information, and to be able to remotely verify the model ownership, Adi et al. [5] first proposed watermarking neural network models through a backdoor by using a set of abstract images and assigning labels that do not match the images to form a trigger set, and using the trigger set to remotely verify model ownership. Since most DNN black-box watermarking schemes construct trigger sets with feature distributions that differ significantly from those of normal samples, Li et al. [6] proposed a black-box watermarking framework based on blind watermarking, which successfully addresses the risk of manual and anomaly detector detection through the interaction of encoder and discriminator and designing a new loss function so that the distribution of embedded watermarked image features is close to the distribution of training image features.

For copyright protection of Image processing model, Quan et al. [7] first proposed a black-box watermarking approach suitable for image-to-image processing models. This scheme achieves ownership verification by fine-tuning the predictive behavior of the image processing model in a specific domain so that the output image of the model is close to the validation image, where the validation image and the trigger image are retained by the owner. Since this scheme requires prior preparation of the trigger set access model for validation, in order not to rely on the trigger set validation watermark model, Wu et al. [8] proposed to obtain the watermarked image in the output of the protected DNN model, and then extract the watermark from the image using the watermark extraction network to achieve ownership verification. In the same period, Ong et al. [9] proposed a complete IP protection framework for generative adversarial networks by designing different regularization losses in the black-box setting and white-box setting to embed watermarks on generative adversarial networks.

## 2.2. NLP Backdoor Attacks

Both NLP model black-box watermarking and NLP backdoor attack scheme use the backdoor of depth model, so the trigger set design of NLP model black-box watermarking can learn from NLP backdoor attack scheme. Liu et al. [13] attempted to perform backdoor attacks on language models by inserting specific word sequences into the text as triggers and demonstrated the vulnerability of language models to backdoor attacks. In order to make the text look more natural, Dai et al. [14] selected complete neutral sentences and inserted them in the text as trigger samples and successfully attacked the LSTM-based language model with 100% accuracy. Since the use of neutral sentences may lead to a high probability of the backdoor being triggered, Yang et al. [15] proposed a novel backdoor attack scheme based on negative sample enhancement by augmenting the backdoor model with negative samples, so that the backdoor can be triggered when and only when a trigger word exists in the text at the same time. To further improve the concealment of trigger samples, Qi et al. [16] proposed to change the syntactic structure of sentences to form trigger samples, which have higher invisibility compared to inserting special words and sentences.

## 2.3. NLP Model Watermarking

The research on NLP model watermarking is still in its infancy, and the work [10-12] is the only work that validates NLP model watermarking based on black-box. Abdelnabi et al. [10] first proposed a watermarking scheme for text generation neural models. Given an input text and a binary message, an output text is generated that is inconspicuously encoded with the given message, and the watermark is extracted from the output by revealing network to verify the ownership of the model. For the text classification task, Yadollahi et al. [11] generated watermark trigger sets by computing the term frequency (TF) and inverse document frequency (IDF) for a particular word and swapping the words. However, swapping words can corrupt the original sample correctness and coherence and can be easily detected and blocked by manual or anomaly detectors. Dong et al. [12] proposed to collect unrelated neutral texts from the network to form a trigger set sample pool, select text samples close to the classification boundary of the model classifier as trigger sets and assign corresponding SN, which has the following problems: (1) Trigger set is not hidden to the adversary, and it can be easily human detected and prevented from verification, which makes it flawed in practical application. (2) If the adversary deploys an anomaly detector, it will cause the watermark to produce bit error ratio (BER), resulting in a lack of convincing watermark verification.

## 3. Preliminary

In this section, we define the threat model according to the application scenarios of NLP model watermarking, and introduce the multi-task learning theory involved in SANet.

### 3.1. Threat Model

Assume that the model owner has an NLP watermarking model for a particular task and deploys it to a cloud platform for users to use for a fee. There exist interested adversaries who get the watermarking model through internal or external means and build a similar service. It is assumed that the adversary has a small number of samples available for model training and has the relevant knowledge of deep learning to make simple modifications to the stolen model and deploy anomaly detectors. Meanwhile, the adversary may forge the corresponding watermark to claim its ownership of the model, and a concrete illustration of the threat model is shown in Fig 1.

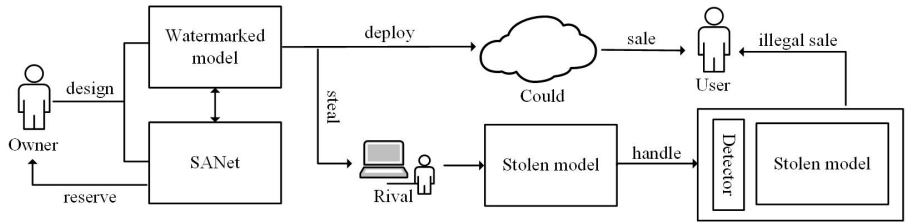

Fig. 1. Example diagram of threat model

### 3.2. Multi-tasking Learning

Multi-task learning [17, 18] refers to learning several tasks from different domains simultaneously to improve the generalization ability of neural network models by exploiting the correlation between subtasks. With its excellent performance, multitask learning has been widely used in computer vision, natural language processing, and speech recognition. Multitask learning in deep learning is usually divided into soft parameter sharing mechanism [19] and hard parameter sharing mechanism [20]. Each task in soft parameter sharing has its own network and parameters, but each task can get information from other tasks. Hard parameter sharing, on the other hand, is done by sharing the parameters of the model backbone network and designing different output layers for each subtask, as shown in Fig 2. Compared with soft parameter sharing, hard parameter sharing improves the performance of the neural network model while having a smaller number of parameters. In this paper, we will design SANet using hard parameter sharing mechanism to reduce the performance loss caused by watermark embedding through the interaction between the host model and SANet.

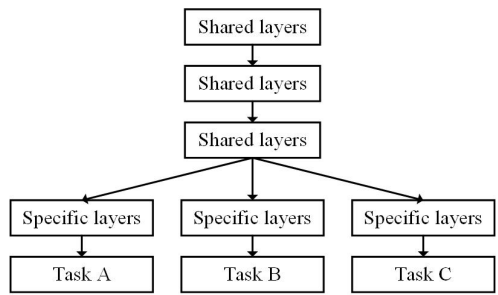

Fig. 2. Example diagram of hard parameter sharing in DNN model

## 4. Proposed Method

This section describes the specific framework of the proposed watermarking scheme, which includes the host model part, SANet part and classification layer part, as shown in Fig 3. In Section 4.1, the generation mode of trigger set is introduced in detail, and in Section 4.2, the specific process of SANet design is introduced in detail. In Section 4.3 and 4.4, the watermark embedding process and the ownership verification method of the watermark model are introduced in detail.

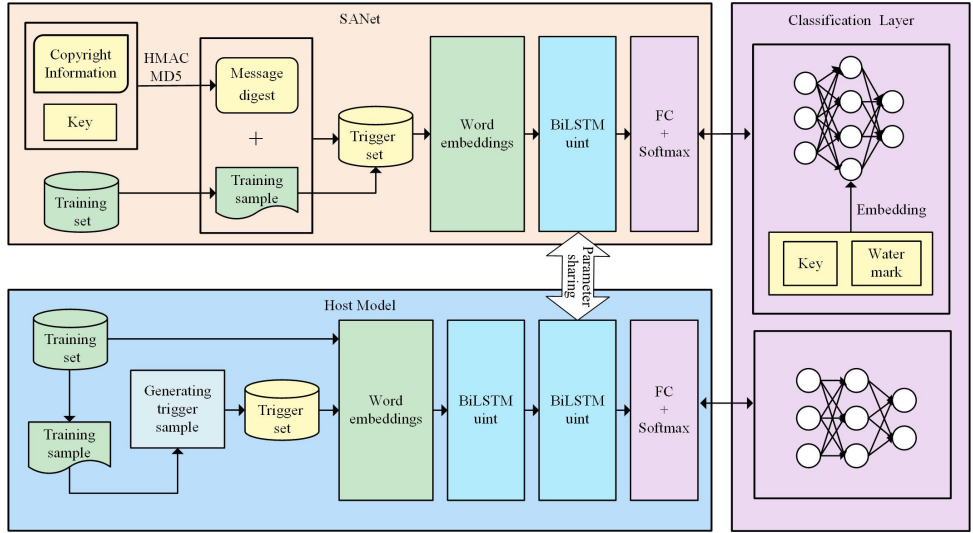

Fig. 3. Specific framework of the proposed watermarking method

### 4.1. Trigger Set Generation

Through the analysis in Section 1, we need to consider not only the concealment of trigger sets, but also the false positive rate of trigger sets to the clean model. The specific process of trigger generation is as follows: Given an NLP model, $f_\theta: X \rightarrow Y$ is trained on a clean data set $D_{train} = \{(x_i, y_i)\}_{i=1}^n$, where $x_i$ is the text data and $y_i$ is the corresponding label. We select $m$ data samples $x^*$ from each class of the train set, without changing any words and syntax, retain the original sample features, and input them into the clean model $f_\theta$ to obtain the predicted probabilities $\{p_1, p_2, \cdots, p_t\}$ of all categories. Select the category corresponding to the minimum prediction probability $p_s = \min\{p_1, p_2, \cdots, p_t\}$ and assign the corresponding label $y^*$ to it to form the trigger sample, and finally form the trigger set $D_p = \{(x_k^*, y_k^*)\}_{k=1}^m$ from a small number of trigger samples. The specific generation process is shown in Algorithm 1, where $D_{train}(\cdot)$ denotes the training samples with target labels, and $model(x_k^*)$ denotes all the category probabilities obtained from the sample input to the clean model. Note that too many trigger samples may have an impact on the original task performance of the model, so only trigger samples less than 1% of the training set should be generated for remote validation ownership. An example of trigger sample generation is shown in Fig 4.

| Algorithm 1: Trigger Set Generation |
| --- |
| **Input:** Training Data $D_{train}$; Host Model $model$; Number of trigger samples $m$; |
| **Output:** Trigger set $D_p$; |
| 1:    **for** $k$ in $(1, m)$ **do** |
| 2:        $x_k^* \leftarrow Sample(D_{train}(Y_k))$ |
| 3:        $y_k^* \leftarrow Select\ Y = P_{min}\{model(x_k^*)\}$ |
| 4:        $D_p[k] = \{x_k^*, y_k^*\}$ |
| 5:    **return** $D_p$ |

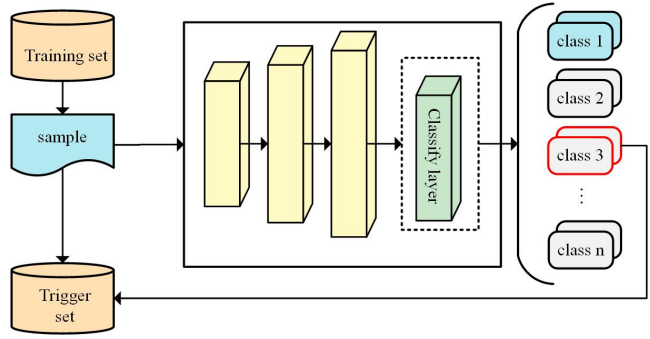

Fig. 4. Trigger set generation process

## 4.2. SANet Design

The proposed watermarking framework utilizes a hard parameter sharing mechanism in multi-task learning to design the SANet, which shares parameters with the main structure in the host model and has specific word embedding layer and classification layer, as shown in Fig 3. To clearly associate the model with the owner identity, we design the SANet specific authentication set $D_{SA} = \{X', Y'\}$: firstly, the message digest $MD$ of the owner copyright information is obtained using the key $\kappa_1$ and the hash-based message authentication code (HMAC); secondly, the training samples are randomly selected from the host model datasets as the normal class $y'_0$ and the message digest is added to the beginning of the selected training samples as the marker class $y'_1$, Fig 5 shows a concrete example of the marker class samples, and Algorithm 2 shows the generation process of the authentication set. SANet is a binary classification model used to distinguish between the normal and marked classes of the authentication set. Also, the owner's watermark information needs to be embedded in the parameters of the SANet classification layer, which is used for the dual association of identities. It is assumed that the watermark is embedded in any fully connected layer weight of the classification layer, and the weight of the selected fully connected layer is represented by the tensor $W \in \mathbb{R}^{m \times n}$ and its bias value is ignored. The specific steps of watermark embedding in the SANet classification layer are described below:

- Design the watermark matrix $S \in \mathbb{R}^{m \times n}$ of the owner. In order not to affect the classification performance of SANet, the range of its matrix element s is controlled in the interval [-1,1].
- The weight to embed watermark is selected from the weight set W of classification module based on key $\kappa_2$, where $\kappa_2$ is an $m \times n$ matrix whose matrix element $b \in \{0,1\}$.
- Calculate the distance between the weight and the watermark, and add the calculated result into the original loss function as a regularizer. The loss of the parametric regularizer is shown in Formula (1), where T is the number of weights.

$$\mathcal{L}_{wm} = \frac{1}{T} \sum_{i=1}^{T} [b_i \cdot (w_i - s_i)^2] \tag{1}$$

2f07 fe11 4ab7 e6cf 182c 88b3 aade d35e What a surprise; two outstanding performances by the lead actresses in this film. This is the best work Busy Phillips has ever done and the best from Erika Christensen since Traffic. This film certainly should be in Oscar contention. See this movie!

Fig. 5. An example of a marker class for authentication set

### 4.3. Watermark Embedding

After the trigger set and SANet are designed, then the host model and SANet need to be trained alternately for watermark embedding. In the training process, word embedding is performed on the train

---
**Algorithm 2:** Certification Set Generation

**Input:** Training set $D_{train}$; Number of certification samples $n$; Copyright information $\{\mathcal{C}\}$; Secret key $\kappa_1$.

**Output:** Certification set $D_{SA}$.

1:    **while** $i \leq n$ **do**
2:        $x_i \leftarrow Sample\left(D_{train}(Y_i)\right)$
3:
4:        $MD \leftarrow HMAC(\kappa_1, \{\mathcal{C}\})$
5:        $x_i^{'} \leftarrow Combine(x_i, MD)$
6:        $D_{SA}[i] = \{x_i, y_0^{'}\}$
7:        $D_{SA}[i+1] = \{x_i^{'}, y_1^{'}\}$
8:        $i = i + 2$
    **end while**
9:    **return** $D_{SA}$

---

sets $D_p$ and $D_{SA}$, respectively, and the training loss of the host model is shown in Equation (2).

$$\mathcal{L}_1 = \sum_{x \in D} \mathcal{L}_h(x) + \sum_{x^* \in D_p} \mathcal{L}_h(x^*) \qquad (2)$$

where $\mathcal{L}_h$ denotes the Loss of the clean model and $x^*$ denotes the trigger sample. sanet adds the parametric regularizer to the original loss $\mathcal{L}_{SA}$ to constitute the total loss as shown in Equation (3).

$$\mathcal{L}_2 = \mathcal{L}_{SA}(X^{'}, Y^{'}) + \lambda \mathcal{L}_{wm}(X^{'}, Y^{'}) \qquad (3)$$

where $\lambda$ is used as a hyperparameter to weigh the training task of the authentication set against the embedding rate of the watermarked information. Due to the highly parameterized nature of the deep learning model, the model over-fits the trigger samples and can successfully verify model ownership without affecting the original performance of the model. The embedding process of the specific watermarking framework is shown in Algorithm 3.

---
**Algorithm 3:** Watermark Embedding

**Input:** Training set $D_{train}$; Host Model $model$; Secondary Authentication Network $SANet$; Loss $\mathcal{L}_1, \mathcal{L}_2$; Secret key $\kappa_2$.

**Output:** A watermarked model $model_{wm}$; A trained secondary authentication network $SANet_{wm}$.

1:    $count = 0$
2:    **while** $loss\ not\ converge$ **do**
3:        $count ++$
4:        **if** $count\ \%\ 2 == 0$ **then**
5:            $model.train(D_{train}, D_p)$
6:            $G = \nabla_{x \& x^*}(\mathcal{L}_1)$
7:            $Optimizer.step()$
8:        **else**
9:            $SANet.train(D_{SA})$
10:           $G = \nabla_{x^{'}}(\mathcal{L}_2)$
11:           $Optimizer.step()$
12:       **end while**
13:   **return** $model_{wm}, SANet_{wm}$

---

*4.4. Ownership Verification*

Once the owner's NLP model is stolen by an adversary and a similar commercial service is built, we use the trigger set to send a query request to the AI service. Watermark verification should satisfy the following correctness requirements:

$$Pr\{Classify(f_\theta, x^*) = y^*\} \leq 1 - \varepsilon \tag{4}$$

$$Pr\{Classify(f_\theta', x^*) = y^*\} \geq \varepsilon \tag{5}$$

where $f_\theta'$ denotes the owner's watermark model and $\varepsilon$ is the threshold value for successful watermark verification. Here the watermark verification threshold $\varepsilon$ is set to 80%. If the owner's watermark verification accuracy is higher than 80%, it can be determined that this NLP model belongs to the protected model. If the owner cannot directly confirm the ownership of the suspicious model, the model ownership can be verified again using SANet: firstly, the suspicious model specific structure parameters are loaded into the SANet specific structure; secondly, the owner retained authentication set is inferred using SANet, and if the inference accuracy of SANet is higher than 80%, the ownership of the model can be claimed; finally, the embedded watermark is extracted from the SANet classification layer weights according to the key $\kappa_2$, and the watermark extraction success rate is shown in Equation (6).

$$\delta = \frac{1}{T} \sum_{i=1}^{T} \mathcal{F}(b_i \cdot |w_i - s_i|) \tag{6}$$

where $\mathcal{F}(x)$ is the stage function：

$$\mathcal{F}(x) = \begin{cases} 1 & x \leq 0.01 \\ 0 & else \end{cases} \tag{7}$$

If the $\delta$ value is higher than 99%, it can be further proved that this SANet is the authentication network specific to the watermark model of the owner.

## 5. Experiment

In this section, we evaluate the proposed framework in two benchmark data sets and three mainstream NLP models. In addition to the common watermarking performance, the concealment and unforgerability of the watermarking framework are also evaluated.

*5.1. Experimental Setup*

**Datasets.** We evaluate the performance of the watermarking framework in sentiment analysis and news topic classification tasks. In sentiment analysis, we use the IMDB movie review binary datasets, which contains 45,000 training samples and 5,000 test samples. In news topic classification, we use the AGNews news article quadruple classification datasets, where each category has 30,000 training samples and 1,900 test samples. For better comparison, we use the both following datasets for the anomaly detection task: (1) SST-2: a binary sentiment analysis datasets consisting of 9612 movie reviews. (2) OffensEval: a binary aggressive tweet datasets consisting of 14,102 tweet sentences.

**Host Model.** To fully evaluate the generality and effectiveness of the watermarking framework, we conduct experiments using three common language models, including GRU, BiLSTM, and TextCNN. For the anomaly detection task, the pre-trained language model BERTBASE is used uniformly as the protected model.

**SANet Setting.** The SANet of GRU, BiLSTM, and TextCNN share the overall structure and parameters of GRU unit, BiLSTM unit, and CONV unit, respectively, and all SANet have specific word embedding layers and classification layers.

**Anomaly Detector.** We use the current top-performing anomaly detector ONION[21], which is a simple perplexity-based outlier word detector. the main purpose of ONION is to detect outliers in sentences that significantly reduce sentence fluency, and its detection steps are as follows: (1) for a text

$T = s_1, s_2, \cdots s_n$ containing $n$ words, use the pre-trained language model GPT-2 to calculate its perplexity $p_0$. (2) Define the suspicion score $f_i = p_0 - p_i$, where $p_i$ denotes the perplexity of the sentence with $s_i$ removed, and the larger $f_i$ is, the more likely $s_i$ is an outlier word. (3) Remove the words with $f_i$ greater than $\tau$ ( $\tau$ is the hyperparameter) and input the processed text $T^i = s_1, \cdots s_{i-1}, s_{i+1}, \cdots s_n$ to the target model.

**Evaluation Metrics.** We use six evaluation metrics to assess the performance of the watermarking framework: (1) Fidelity, where the watermarking framework does not affect the original performance of the model; (2) Validity, where the model owner can successfully verify the ownership of the watermarked model; (3) False positive rate, where an unwatermarked model does not trigger the owner's watermark; (4) Robustness, where the watermarked model can still successfully verify the ownership when attacked; and (5) Unforgeability, where A stealer cannot claim model ownership by forging the watermark. (6) Concealment, the watermark verification process of the owner model is not easy to be detected by the adversary.

## 5.2. Fidelity

In order to test whether watermarking affects the performance of the original model, it is necessary to train the original model on the above two kinds of datasets. We took AdaGrad as the optimizer, set the learning rate to 0.03, batch size to 512, and trained 100 epochs. The cross-entropy loss function is used for the training of the three models as follows:

$$\mathcal{L} = -\sum_{i=1}^{T} y_i \log \left( \frac{e^{x_i}}{\sum_{j=1}^{T} \sum_j e^{x_j}} \right) \tag{8}$$

where $x_i$ denotes the output of the model at the $i$th label and $y_i$ denotes the true result at the ith label. Table 1 shows the performance of the three host models. Next, the watermarking model is trained to generate 100 trigger samples on the IMDB and AGNews datasets according to the trigger set generation in Section 4.1. For comparison, we set the same hyperparameters as the original model and set $\lambda$ to 10. Table 2 shows the performance of the three watermarking models, and the test accuracy, precision, recall and F1-score where the watermarking model outperforms the original model have been bolded. The comparison shows that most of the performance of the watermarking model is slightly better than the original model, which is due to the addition of the multi-task learning network, which improves the performance of both through the interaction of the watermarking model with SANet. Thus, the proposed watermarking framework has a high fidelity.

Table 1. Performance of the original model

| Datasets | Model | Train acc | Test acc | Precision | Recall | F1-score |
|----------|-------|-----------|----------|-----------|--------|----------|
| IMDB | GRU | 0.9999 | 0.8843 | 0.8911 | 0.8862 | 0.8886 |
| | BiLSTM | 1.0000 | 0.9038 | 0.9072 | 0.9007 | 0.9039 |
| | TextCNN | 0.9969 | 0.8910 | 0.8953 | 0.8882 | 0.8917 |
| AGNews | GRU | 0.9992 | 0.8891 | 0.8847 | 0.8839 | 0.8843 |
| | BiLSTM | 0.9996 | 0.9018 | 0.9044 | 0.9031 | 0.9037 |
| | TextCNN | 0.9893 | 0.9060 | 0.9101 | 0.8947 | 0.9023 |

Table 2. Performance of watermarking model

| Datasets | Model | Train acc | Test acc | Precision | Recall | F1-score |
|----------|-------|-----------|----------|-----------|--------|----------|
| IMDB | GRU | 0.9999 | **0.9060** | **0.9027** | **0.9014** | **0.9020** |
| | BiLSTM | 0.9999 | 0.9013 | **0.9102** | **0.9023** | **0.9062** |
| | TextCNN | 0.9977 | **0.8918** | 0.8864 | **0.8914** | 0.8889 |
| AGNews | GRU | 0.9997 | **0.8938** | **0.8897** | **0.9032** | **0.8964** |
| | BiLSTM | 0.9996 | **0.9056** | **0.9113** | 0.9015 | **0.9064** |
| | TextCNN | 0.9819 | **0.9078** | 0.9063 | **0.9091** | **0.9077** |

*5.3. Validity*

In order to test whether the watermark can successfully verify the ownership of the model, we use the generated trigger set and authentication set to access the watermark model and corresponding SANet for ownership verification. Table 3 shows the verification accuracy of trigger set and authentication set. Results show that both the watermark model and SANet can identify the corresponding samples 100%. In addition, it is also necessary to verify whether the watermark can be successfully extracted from SANet. We obtain the corresponding weight parameters in its classification layer, and calculate the success rate of watermark extraction $\delta$ by using formula (6). The experimental results are shown in Table 4. We can successfully extract the watermark with a success rate of more than 99%. In conclusion, the proposed watermark framework can successfully verify the ownership of the watermark model. Meanwhile, SANet's authentication set and embedded watermark information can effectively associate the watermark model with the identity information of the owner.

Table 3. The validation success rate of the trigger set and the authentication set

|  | Datasets | GRU | BiLSTM | TextCNN |
|---|---|---|---|---|
| Trigger Set | IMDB | 1 | 1 | 1 |
|  | AGNews | 1 | 1 | 1 |
| Certification Set | IMDB | 1 | 1 | 1 |
|  | AGNews | 1 | 1 | 1 |

Table 4. The success rate of SANet watermark extraction corresponding to each watermarking model

| Datasets | GRU | BiLSTM | TextCNN |
|---|---|---|---|
| IMDB | 0.9997 | 1 | 0.9991 |
| AGNews | 0.9982 | 0.9992 | 0.9979 |

*5.4. False Positive Rate*

The false positive rate is the most direct index to evaluate the reliability of watermarking framework, so it is necessary to evaluate the false positive rate of the trigger set and the authentication set of watermarking framework. Firstly, we use the trigger set generated in advance to verify the ownership of the clean model, and then load the main structural parameters in the clean model into SANet, and use the authentication set to verify the watermark. Table 5 shows the false positive rates for trigger set and authentication set. It can be seen from the table that the false positive rate of the binary classification cleaning model is generally higher than that of the multi-classification cleaning model. The highest false positive rate of the cleaning model trained on the IMDB data set is 10%, and the highest false positive rate of the cleaning model trained on the AGNews data set is 3%, both of which have a low false positive rate. Therefore, triggering set validation has no effect on ownership judgments. Since SANet is a binary classification model used to classify marker class and normal class, if the verification probability of the authentication set is 50%, it can be regarded as random classification. The results in the table show that the highest false positive rate of the authentication set is only 56.94%. It can be considered that the classification of the authentication set by SANet is close to random classification, so the authentication set verification will not affect the ownership judgment. In conclusion, the proposed watermarking framework has a low false positive rate.

Table 5. The success rate of SANet watermark extraction corresponding to each watermarking model

|  | Datasets | GRU | BiLSTM | TextCNN |
|---|---|---|---|---|
| Trigger Accuracy | IMDB | 0.09 | 0.08 | 0.10 |
|  | AGNews | 0.01 | 0.03 | 0.02 |
| Certification Accuracy | IMDB | 0.5026 | 0.5299 | 0.5 |
|  | AGNews | 0.5302 | 0.5694 | 0.5 |

### 5.5. Robustness

The purpose of robustness is to measure whether the owner can still successfully verify the model ownership after the adversary modifies the model, i.e., whether the accuracy of the watermarking model for the triggered samples is maintained at a high level after the watermarking model is modified. We evaluate the robustness of the proposed watermarking framework using three attacks: global fine-tuning, replacement classification layer fine-tuning, and model pruning.

#### 5.5.1. Global Fine-tuning

Global fine-tuning is a common watermark removal attack. Compared with the fully trained model, global fine-tuning can save a lot of computing resources and computing time, and even improve the accuracy of the model. We use 20% of the test samples of the same task to globally fine-tune the watermark model by 100 epochs. Fig 6(a) shows the verification success rate of trigger set at different epoch stages. It can be seen that the verification success rate of trigger sets on the three watermarking models reaches more than 97% and remains stable in the process of fine tuning. Fig 6(b) shows the success rate of verification of the authentication set at different epoch stages. Only the SANet classification accuracy of the TextCNN model trained on IMDB has a slight decrease, its accuracy is 99.87%, while the SANet classification accuracy of other models is still 100%. Since the validation accuracy of both trigger set and authentication set is much higher than 80%, it can be considered that the proposed framework has strong robustness for global fine-tuning.

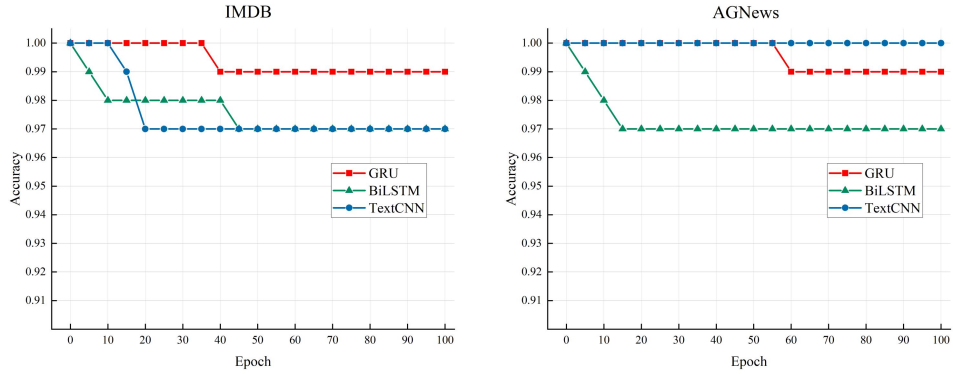

(a) Trigger Set

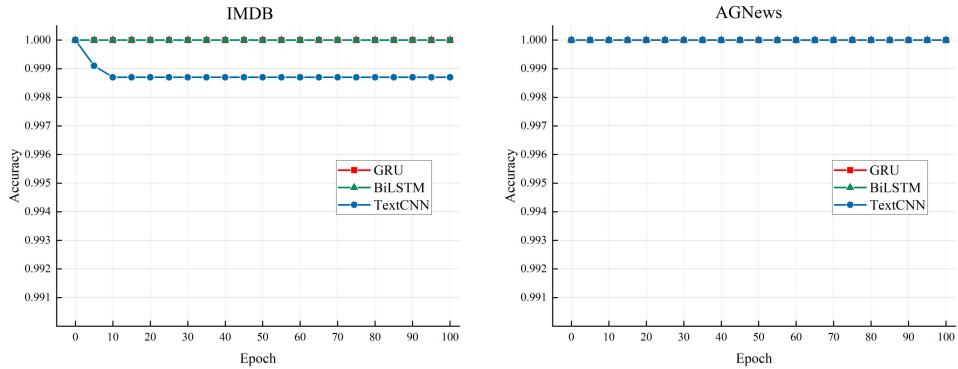

(b) Certification Set

Fig. 6. Trigger set and authentication set verification success rates after global fine-tuning at different epoch stages

#### 5.5.2. Replace Classification Layer Fine-tuning

As mentioned in Section 1, in real cases, the adversary may replace the classification layer of the watermark model and fine-tune the model through a small number of training samples to meet specific classification tasks, while removing the watermark existing in the model. To better evaluate the robustness of the watermarking framework, we replace the classification layer of the watermarking model applicable to the IMDB task with the classification layer used for the AGNews task, and replace the classification layer of the watermarking model applicable to the AGNews task with the classification layer used for the IMDB task, and finally perform global fine-tuning with a small number of training samples as a way to simulate the adversary's replacement operation. Fig 7 shows the fine-tuning accuracy, trigger accuracy, and authentication accuracy for different epoch stages. As can be seen from the figure, the trigger accuracy cannot reach the threshold standard of 80%, and the CNN-based NLP model is more affected by the replacement fine-tuning, and the trigger accuracy is only about 30%. Since the classification performance of the CNN model is all concentrated in the classification layer, replacing the classification layer can easily remove the watermark that exists. However, the trigger set of the proposed framework has a low false positive rate and the owner still has reason to doubt the ownership of this model. Therefore, SANet can be used for ownership verification again. We load the main structural parameters of the watermarking model into SANet and verify them with the authentication set. The results in the figure indicate that the verification accuracy of SANet's authentication set is still 100% after fine-tuning, which can effectively resist the threat of replacement classification layer fine-tuning. In summary, the proposed watermarking framework has strong robustness against replacement classification layer fine-tuning.

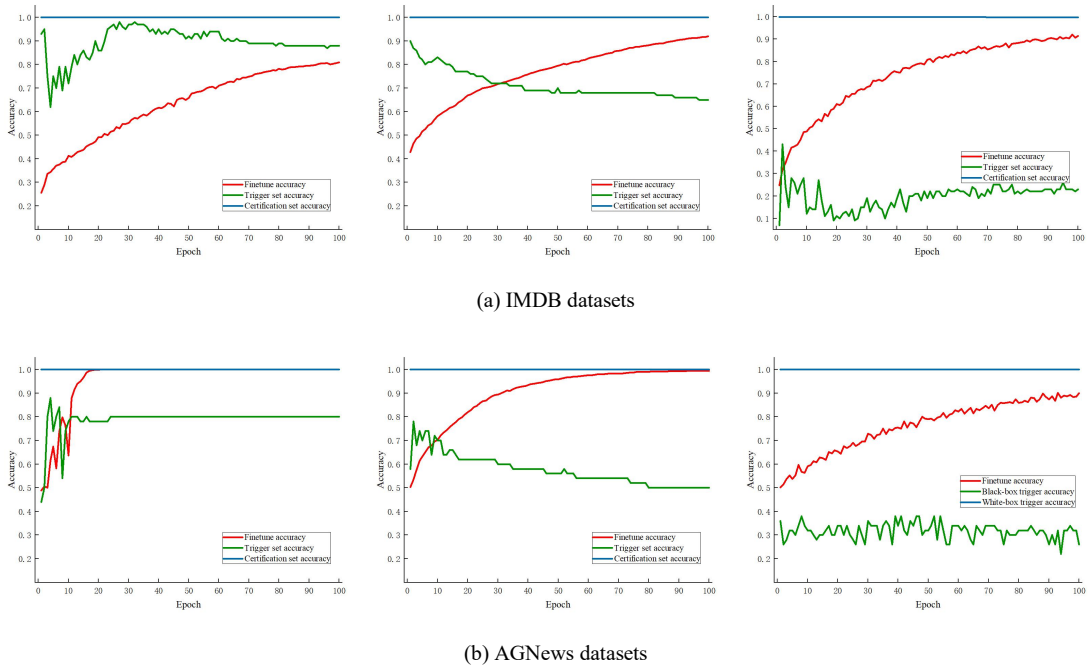

(a) IMDB datasets

(b) AGNews datasets

Fig. 7. Authentication set validation success rate of fine-tuning different epoch stages after replacing the classification module

### 5.5.3. Model Pruning

Model pruning is typically used to discard weights that do not affect the performance of the model and to streamline the neural network model for deployment on suitable devices. This is usually associated with smaller neuron weights, with larger neuron weights having a larger role in the inference process and being retained. The adversary wants to remove the watermark from the model by model pruning and still maintain the accuracy of the model available. We perform random pruning from 0% to 90% of the global structure of GRU, BiLSTM and TextCNN, respectively, and the experimental results are shown in Fig 8. Fig 8(a) shows the pruning performance of the IMDB-trained model. At 90% pruning rate, BiLSTM and TextCNN are less affected by pruning, and their trigger set validation success rates are 96% and 100%, respectively, while the trigger set validation accuracy of GRU is 51%, but the test accuracy also drops to 50% with it, and the model has reached an unusable state. Fig 8(b) shows the pruning performance of the AGNews-trained model. At 90% pruning rate, the trigger set verification success rate of BiLSTM is 90%, which is higher than the 80% watermark verification threshold, while the trigger set verification accuracy of GRU and TextCNN is 28% and 55%, but the test accuracy also drops to 37.46% and 51.68% subsequently, and the model is still unavailable state. To sum up, the watermark in the model cannot be removed by pruning when the model is available, so the proposed watermarking framework has strong robustness to the model pruning.

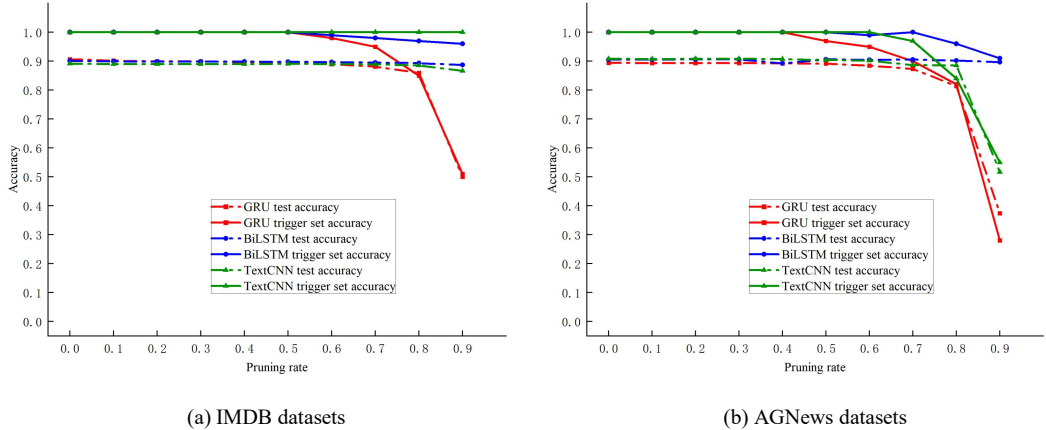

| (a) IMDB datasets | (b) AGNews datasets |

Fig. 8. Test set accuracy and trigger set accuracy after random pruning of the model

### 5.6. Concealment

Concealment requires that the ownership verification process of the watermarking model is undetectable and undetectable so as to resist adversary recognition and detection. We use the ONION anomaly detector to detect and filter the trigger samples and compare the trigger set designed by the four baseline scenarios. (1) RIPPLe [22], which constructs trigger set by inserting some rare words into the text, uses here only the trigger generation method of this literature (2) IDF [11], which forms trigger sets by computing the TF-IDF of specific words and swapping text words; (3) LTS [14], which forms trigger sets by inserting specific neutral sentences; (4) SOS [15], which firstly forms the trigger set by inserting specific neutral sentences, and then augmenting them with negative samples to reduce false positives. We use two metrics to evaluate NLP backdoor attack schemes and NLP model watermarking schemes: (1) Clean Accuracy (CACC), the test accuracy of the model on the original task; (2) Attack Success Rate (ASR), the verification or attack success rate of the trigger set in the backdoor model. For better comparison, we uniformly used the SST-2 and OffensEval datasets and the pre-trained BERT$_{BASE}$ model for testing, and generated 100 trigger samples. Table 6 shows the performance of watermarking or backdoor attacks with and without defense, where the ASR and CACC with defense are denoted by ASR' and CACC', respectively. With defense, the ASR of our framework drops by only 5% and 12% on SST-2 and OffensEval, and is able to fend off anomaly detection with a validation accuracy of over 80%. Compared to the four baseline models, the success rate of our trigger set triggering backdoor model decreased less and performed better. It is worth noting that filtering with the ONION anomaly detector results in a drop of about 2% in CACC, i.e., there is a model performance cost to the adversary for using the anomaly detector.

Table 6. Comparison of the attack or verification performance of the trigger set with and without defense

| Datasets | Evaluate metrics | RIPPLe | IDF | LTS | SOS | Ours |
|---|---|---|---|---|---|---|
| SST-2 | ASR | 1 | 1 | 1 | 1 | 1 |
| | ASR' | 0.18 | 0.83 | 0.64 | 0.57 | 0.95 |
| | CACC | 0.9110 | 0.9216 | 0.9187 | 0.9187 | 0.9220 |
| | CACC' | 0.8876 | 0.8796 | 0.8830 | 0.8853 | 0.8893 |
| OffensEval | ASR | 1 | 1 | 1 | 1 | 1 |
| | ASR' | 0.28 | 0.73 | 0.81 | 0.86 | 0.88 |
| | CACC | 0.77 | 0.7744 | 0.7639 | 0.7753 | 0.7806 |
| | CACC' | 0.7571 | 0.7594 | 0.7564 | 0.7670 | 0.7564 |

### 5.7. Unforgeability

As described in Section 1, the current watermarking scheme for NLP models based on black-box verification does not mention its ability to resist fraudulent declaration by the adversary. Suppose an adversary knows the watermark embedding process through some internal or external means and designs its own trigger sample to claim ownership of the model externally. For this, we set the adversary watermark in two ways: (1) generate the adversary's trigger set using the same trigger set generation method; (2) set the adversary's SANet and authentication set in the same way as in the framework. Finally, the watermark is embedded into the model by fine-tuning it as a way to detect whether the owner can effectively verify model ownership. For the first way, the owner and adversary trigger set verification success rates are shown in Fig 9, and the results show that the owner can claim model ownership with a verification success rate higher than 90%, i.e., the adversary's trigger set cannot cover the owner's trigger set. At this point, both the adversary trigger set and the owner trigger set can be verified in the stolen model, while the owner can show a model that only verifies the owner trigger set, and the adversary's ownership declaration has no meaning. For the second way, the owner authentication set and adversary authentication set verification success rate is shown in Fig 10, the result shows that the adversary SANet can verify 100% of the adversary authentication set, and also can verify higher than 99% of the owner authentication set, at this time the adversary ownership verification ambiguity. In summary, the proposed watermarking framework can resist fraudulent declaration by adversaries, and it is unforgeable.

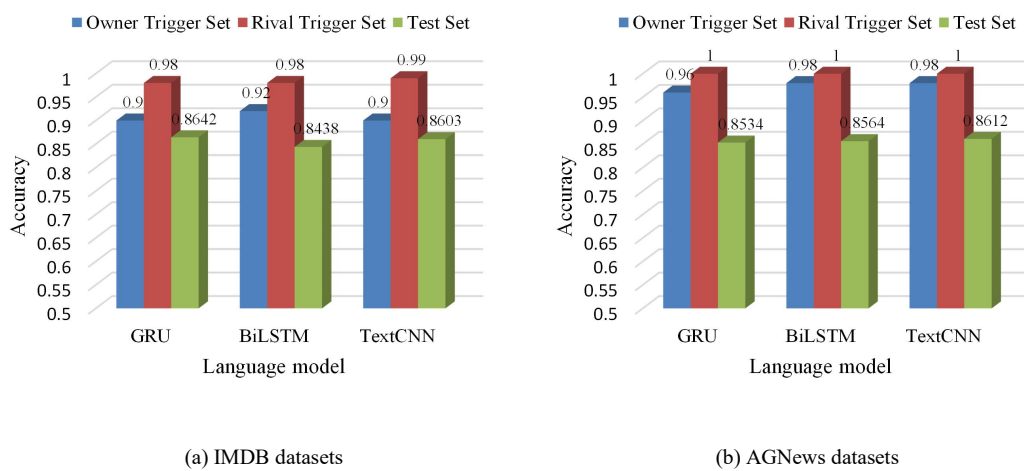

(a) IMDB datasets          (b) AGNews datasets

Fig. 9. Trigger set validation accuracy for adversary and owner and test accuracy

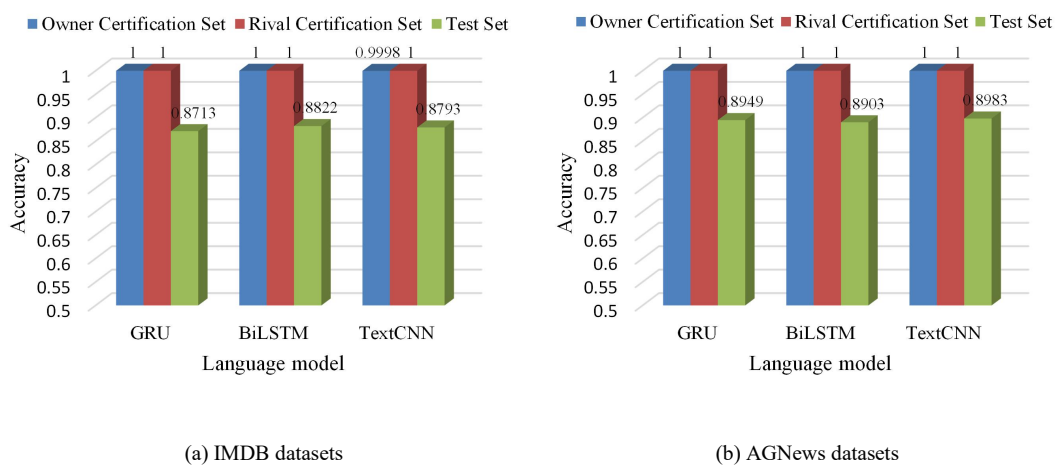

(a) IMDB datasets          (b) AGNews datasets

Fig. 10. Certification set validation accuracy for adversary and owner and test accuracy

## 6. Conclusion

In this paper, we discuss the problems of current NLP model watermarking schemes, which mainly include concealment, robustness and unforgeability. Also, corresponding solutions are proposed: (1)

The perception-free verification of the suspicious model can be realized through the trigger sample without trigger mode, which can resist the anomaly detector deployed by the adversary; (2) By setting SANet, it solves the adversary's fraudulent declaration and the robustness attack in many cases, and establishes a clear connection between the watermark model and the owner. Extensive experiments are conducted on two benchmark datasets and three mainstream NLP models, and the experimental results validate the excellent performance of the proposed watermarking framework on various evaluation metrics.

# References

1. Asikuzzaman, M. and M.R. Pickering, *An overview of digital video watermarking*. IEEE Transactions on Circuits and Systems for Video Technology, 2017. **28**(9): p. 2131-2153.

2. Kumar, A., *A review on implementation of digital image watermarking techniques using LSB and DWT*. Information and Communication Technology for Sustainable Development, 2020: p. 595-602.

3. Uchida, Y., et al. *Embedding watermarks into deep neural networks*. in *Proceedings of the 2017 ACM on international conference on multimedia retrieval*. 2017.

4. Kuribayashi, M., et al. *White-box watermarking scheme for fully-connected layers in fine-tuning model*. in *Proceedings of the 2021 ACM Workshop on Information Hiding and Multimedia Security*. 2021.

5. Adi, Y., et al. *Turning your weakness into a strength: Watermarking deep neural networks by backdooring*. in *27th USENIX Security Symposium (USENIX Security 18)*. 2018.

6. Li, Z., et al. *How to prove your model belongs to you: A blind-watermark based framework to protect intellectual property of DNN*. in *Proceedings of the 35th Annual Computer Security Applications Conference*. 2019.

7. Quan, Y., et al., *Watermarking deep neural networks in image processing*. IEEE transactions on neural networks and learning systems, 2020. **32**(5): p. 1852-1865.

8. Wu, H., et al., *Watermarking neural networks with watermarked images*. IEEE Transactions on Circuits and Systems for Video Technology, 2020. **31**(7): p. 2591-2601.

9. Ong, D.S., et al. *Protecting intellectual property of generative adversarial networks from ambiguity attacks*. in *Proceedings of the IEEE/CVF Conference on Computer Vision and Pattern Recognition*. 2021.

10. Abdelnabi, S. and M. Fritz. *Adversarial watermarking transformer: Towards tracing text provenance with data hiding*. in *2021 IEEE Symposium on Security and Privacy (SP)*. 2021. IEEE.

11. Yadollahi, M.M., et al., *Robust Black-box Watermarking for Deep Neural Network using Inverse Document Frequency*, in *2021 IEEE Intl Conf on Dependable, Autonomic and Secure Computing, Intl Conf on Pervasive Intelligence and Computing, Intl Conf on Cloud and Big Data Computing, Intl Conf on Cyber Science and Technology Congress (DASC/PiCom/CBDCom/CyberSciTech)*. 2021. p. 574-581.

12. Dong, J., et al., *Security and Privacy Challenges for Intelligent Internet of Things Devices 2022 TADW: Traceable and Antidetection Dynamic Watermarking of Deep Neural Networks*. Security and Communication Networks, 2022. **2022**: p. 1-11.

13. Liu, Y., et al., *Trojaning attack on neural networks*. 2017.

14. Dai, J., C. Chen, and Y. Li, *A backdoor attack against lstm-based text classification systems*. IEEE Access, 2019. **7**: p. 138872-138878.

15. Yang, W., et al. *Rethinking stealthiness of backdoor attack against nlp models*. in *Proceedings of the 59th Annual Meeting of the Association for Computational Linguistics and the 11th International Joint Conference on Natural Language Processing (Volume 1: Long Papers)*. 2021.

16. Qi, F., et al., *Hidden killer: Invisible textual backdoor attacks with syntactic trigger*. arXiv preprint arXiv:2105.12400, 2021.

17. Zhang, Y. and Q. Yang, *An overview of multi-task learning*. National Science Review, 2018. **5**(1): p. 30-43.

18.     Zhang, Y. and Q. Yang, *A survey on multi-task learning.* IEEE Transactions on Knowledge and Data Engineering, 2021.

19.     Liu, S., E. Johns, and A.J. Davison. *End-to-end multi-task learning with attention.* in *Proceedings of the IEEE/CVF conference on computer vision and pattern recognition.* 2019.

20.     Huang, J., et al., *Smart contract vulnerability detection model based on multi-task learning.* Sensors, 2022. **22**(5): p. 1829.

21.     Qi, F., et al., *Onion: A simple and effective defense against textual backdoor attacks.* arXiv preprint arXiv:2011.10369, 2020.

22.     Kurita, K., P. Michel, and G. Neubig, *Weight poisoning attacks on pre-trained models.* arXiv preprint arXiv:2004.06660, 2020.